\documentclass[10pt,conference]{IEEEtran}
\ifCLASSINFOpdf

\else

\fi

\ifCLASSOPTIONcompsoc
\usepackage[caption=false, font=normalsize, labelfont=sf, textfont=sf]{subfig}
\else
\usepackage[caption=false, font=normalsize]{subfig}
\fi
\usepackage{lipsum}%

\usepackage{balance}
\usepackage{multicol}   
\usepackage{cite}
\usepackage{gensymb}
\usepackage{multirow}
\usepackage{graphics}  
\usepackage{epsfig} 
\usepackage{graphicx}
\usepackage{epstopdf}
\usepackage{textcomp}
\usepackage{amsmath}
\usepackage{mathtools}
\usepackage{filecontents}
\usepackage{lipsum,color}
\usepackage{mathtools,tikz,caption}
\usepackage{amssymb}
\usepackage{amsfonts}
\usepackage{float,stfloats,blindtext}
\usepackage{mathtools}
\usepackage{hyperref}
\usepackage{amsmath}
\usepackage{algorithm,algorithmic}
\usepackage{upgreek}
\usepackage{epstopdf}
\usepackage{placeins}
\definecolor{royalblue}{RGB}{65,55,255}
\begin{document}
%
\title{\fontsize{20}{23}\selectfont Energy-Efficient Precoding for Dense VCSEL-Based OWC Systems Under a Cooperative Broadcast Model}

\author{
	\IEEEauthorblockN{Hossein Safi\IEEEauthorrefmark{2}, Asim Ihsan\IEEEauthorrefmark{2}, Hossien B. Eldeeb\IEEEauthorrefmark{2}, Bastien Béchadergue\IEEEauthorrefmark{3}, Iman Tavakkolnia\IEEEauthorrefmark{2}, and Harald Haas\IEEEauthorrefmark{2}}
	\IEEEauthorblockA{\IEEEauthorrefmark{2}LiFi Research and Development Centre, Department of Engineering, Cambridge University, UK \IEEEauthorblockA{\IEEEauthorrefmark{3}  Laboratoire d’Ingénierie des Systèmes de Versailles, Université Paris-Saclay, France}
			Emails: \{hs905, ai422, hbe25, it360, huh21\}@cam.ac.uk, bastien.bechadergue@uvsq.fr} \vspace{-5mm}
}
\maketitle

\begin{abstract}
As 6G and beyond aim for sustainable, high-capacity wireless connectivity, optical wireless communication (OWC) has emerged as a compelling solution. Recent advances in vertical-cavity surface-emitting laser (VCSEL) arrays have significantly enhanced OWC performance, enabling high-speed, low-power data transmission. However, dense VCSEL deployments introduce challenges related to interference and energy efficiency (EE). This paper proposes a scalable precoding framework for EE maximization in fully cooperative VCSEL-based OWC broadcast systems. We formulate a non-convex optimization problem to design the precoding matrix under practical optical constraints while guaranteeing minimum user rates. To solve this, we apply Dinkelbach’s method to handle the fractional objective and the inner approximation technique to iteratively convexify and solve the problem. Simulation results show that our approach consistently outperforms regularized zero-forcing in terms of EE, particularly in large-scale deployments, demonstrating its potential for next-generation sustainable dense OWC networks.
\end{abstract}


\section{Introduction}
The rising demand for ultra-reliable, low-latency, and high-throughput wireless communication, driven by applications like autonomous robotics, the metaverse, and the industrial IoT, highlights the limitations of traditional RF systems \cite{6G-1, 6G-NoN2}. Optical wireless communication (OWC) offers a compelling alternative, utilizing the unlicensed optical spectrum to achieve gigabit-per-second speeds with low latency and high energy efficiency \cite{LiFi2}. Additionally, OWC provides high spatial reuse, resistance to electromagnetic interference, and enhanced physical-layer security due to its directional light propagation \cite{fso6g}. Early indoor OWC systems relied on LEDs, but their limited modulation bandwidth and wide beam profiles restricted data rates and spatial resolution. VCSELs have recently emerged as a superior alternative, offering compact size, low power use, and scalability into dense 2D arrays \cite{LiFi2, VCSEL1, VCSEL2}. Their narrow, focused beams enable multi-gigabit speeds over longer distances, making them ideal for high-capacity links in dense user environments \cite{VCSEL-ee}.

To ensure full spatial coverage and high capacity indoors, VCSEL-based OWC systems use densely packed laser arrays \cite{VCSEL3, VCSEL4}. However, this setup creates a multi-user broadcast channel, where overlapping signals cause multi-user interference (MUI) that degrades performance \cite{MUI-1}. Receiver-side solutions like NOMA, interference cancellation, and angle-diversity detection have been explored \cite{NOMA-1, angle-1, MIMO2}, but they often demand complex processing and precise coordination, which are impractical in dynamic environments \cite{NOMA-2}. A more scalable approach is transmitter-side MUI mitigation, with precoding, using channel state information to pre-process signals, emerging as a powerful method for enabling simultaneous multi-user communication \cite{MUI-1, MUI-2, MUI-3, MUI-4, MUI-5, MUI-6}.

In this study, we explore a system model utilizing a fully cooperative dense VCSEL array, functioning similarly to a coordinated multi-point system in RF \cite{MUI-1, Comp}. Here, the entire array of VCSELs acts as a single, coordinated transmitter, jointly serving all users within the environment. This cooperative broadcast approach allows for centralized and sophisticated joint signal processing across all transmitting elements. Technologies such as tunable micro-lens arrays or reconfigurable intelligent surfaces can further enhance this model by dynamically steering and shaping the collective beam pattern for optimal coverage and signal delivery \cite{lidar2}.  Consequently, with signals from multiple VCSELs superimposed at each user's photodetector (PD), linear precoding techniques similar to zero-forcing (ZF) or regularized ZF (RZF) can be effectively applied at the transmitter to spatially separate these signals and suppress interference \cite{MUI-2, MUI-7, MUI-8}.

While precoding has been widely studied for MUI management in OWC systems \cite{MUI-1, MUI-2, MUI-3, MUI-4, MUI-5, MUI-6, MUI-7, MUI-8}, existing research has largely overlooked the challenge of optimizing EE in dense, fully cooperative VCSEL-based broadcast networks. In particular, there is a lack of frameworks that jointly address EE maximization and interference suppression under the unique constraints of large-scale VCSEL arrays operating as centralized transmitters. Also, prior studies on EE often focused on LED-based visible light communication \cite{EE1, EE2, EE3}, whose system characteristics (e.g., broader beams, lower speeds, different constraints) differ from infrared VCSEL systems. More recently, a VCSEL-specific work has explored EE, using techniques like RZF precoding \cite{Ncube}. However, conventional linear precoders like ZF or RZF, while effective for interference suppression, can be highly power-inefficient, especially as the number of transmitters and users grows in dense deployments. Furthermore, many existing approaches do not fully leverage the potential of system-wide cooperation across the entire array or specifically target the maximization of EE under the strict non-negativity, amplitude, and power constraints inherent to OWC systems \cite{MUI-1}.

Thus, our key contribution in this work lies in the design of a novel and scalable precoding and optimization framework for maximizing EE in dense VCSEL-based OWC systems modeled as fully cooperative broadcast networks. We consider a system-wide joint transmission model in which all VCSELs in the array cooperatively serve all users. Under this broadcast formulation, we develop an energy-efficient joint precoding and power control strategy that explicitly accounts for the structure of the broadcast interference channel and the constraints of optical intensity modulation. The resulting EE maximization problem is non-convex due to its fractional objective and non-linear constraints. To tackle this, we adopt a two-stage solution: (i) Dinkelbach’s method transforms the fractional objective into a tractable subtractive form, and (ii) the inner approximation (IA) technique is applied to iteratively solve a sequence of convex subproblems. 

In contrast to conventional RZF, which typically minimizes interference without considering overall system efficiency or fairness, our approach jointly optimizes the precoding matrix to both maximize EE and ensure that all users achieve a guaranteed minimum data rate. This quality of service aware design makes the system more robust to user heterogeneity and network density. Moreover, our method inherently balances power consumption through constrained precoding without requiring explicit power control while respecting amplitude and non-negativity constraints. Simulation results demonstrate that our framework significantly outperforms RZF in terms of EE and fairness, particularly in high-density deployments, establishing a scalable and practical solution for next-generation OWC broadcast systems.

The rest of this paper is organized as follows. Section \ref{II} details the system model for the dense, cooperative VCSEL broadcast channel. Section \ref{eeanalysis} provides EE analysis. Section \ref{III} presents the EE maximization problem formulation and our proposed iterative solution based on Dinkelbach's method and IA. Section \ref{IV} contains simulation results evaluating the performance of our approach against benchmarks. Finally, Section \ref{V} concludes the paper.

\section{System Model}
\label{II}
\subsection{Link Configuration}
We consider a three-dimensional indoor environment with dimensions \( W \times L \times H \), where \( W \) is the room width, \( L \) the length, and \( H \) the height \cite{VCSEL3}. The system consists of \( K \) users and \( V \) VCSEL elements arranged in a dense two-dimensional array mounted on the ceiling. Each user is indexed by \( k \in \{1, 2, \dots, K\} \), and each VCSEL by \( v \in \{1, 2, \dots, V\} \). The users are randomly distributed within the room, with spatial coordinates given by \( [x_k, y_k, z_k] \). The horizontal coordinates follow uniform distributions, i.e., \( x_k \sim \mathcal{U}(0, W) \) and \( y_k \sim \mathcal{U}(0, L) \), while the vertical coordinate is chosen from a set of discrete heights: \( z_k \in \{0.5\,\text{m}, 1\,\text{m}, 1.5\,\text{m}, 2\,\text{m} \} \), representing different user device elevations. We assume that all VCSELs operate cooperatively in a broadcast mode, jointly transmitting to all users within the space. This configuration results in significant spatial beam overlap, necessitating joint precoding across the full array to mitigate MUI. The communication model incorporates line-of-sight (LoS) optical channel characteristics and assumes additive white Gaussian noise at the receiver \cite{MUI-2, MUI-4}.

The VCSELs are arranged in a grid pattern (array) on the ceiling \cite{LiFi2, VCSEL3, safi20253d}, with the $N_{\text{r}}$  number of rows and $N_{\text{c}}$  columns. The pitch distance between adjacent VCSELs is fixed at $\textrm{pi}_x = \textrm{pi}_y$ in both the $x-$ and $y-$directions. The VCSEL grid is centered at the ceiling, and all VCSELs are positioned at the height $H_{\text{c}}$ from the floor. The position of each VCSEL in the array is determined by the following equation 
\begin{flalign}
	(x_{{v}}, y_{{v}}, z_{{v}}) = \left( 
	\begin{array}{l}
		(i - 1) \cdot \textrm{pi}_x - \frac{(N_{\text{r}} - 1) \cdot \textrm{pi}_x}{2} + \frac{W}{2}, \\
		(j - 1) \cdot \textrm{pi}_y - \frac{(N_{\text{c}} - 1) \cdot \textrm{pi}_y}{2} + \frac{L}{2}, \\
		H_{\text{c}}
	\end{array}
	\right),
	\label{eq:vcsel_position}
\end{flalign}
where $i$ and $j$ are the row and column indices.


The optical channel in the system is modeled based on the propagation characteristics of light from the VCSELs to the users, considering the Gaussian beam model \cite{VCSEL3, VCSEL4}. The VCSELs emit light with a beam waist $w_0$ and wavelength $\lambda$, both of which influence light propagation through space. The Rayleigh range, which characterizes the beam's divergence, is given by
\begin{equation}
z_R = \frac{\pi w_0^2}{\lambda}.
\label{Rayleigh range}
\end{equation}
The Euclidean distance $d_{k,v}$ between the user $k$ and the VCSEL $v$, which is required to compute the channel gain, is calculated using the three-dimensional distance formula
\begin{equation}
d_{k,v} = \sqrt{(x_{v} - x_{k})^2 + (y_{v} - y_{k})^2 + (z_{v} - z_{k})^2}.
\label{Rayleigh range}
\end{equation}
To model the optical propagation more accurately, the total distance 
$d_{k,v}$ is split into axial and radial components. The axial distance $d_0$
is the distance between the VCSEL and the user along the z-axis, representing the vertical separation. Hence, it is calculated as
\begin{equation}
d_0^{k,v} = |z_{{v}} - z_{{k}}|
,
\label{axial distance}
\end{equation}
where $ z_{{v}}$ and $z_{{k}}$ are the height coordinates of the VCSEL $v$ and the user $k$, respectively. The radial distance, $r_0$, represents the horizontal distance between the VCSEL and user, and it is given by
\begin{equation}
r_0^{k,v} = \sqrt{(x_{{v}} - x_{{k}})^2 + (y_{{v}} - y_{{k}})^2}.
\label{radial distance}
\end{equation}
These distances are crucial, as the optical beam properties such as beam radius and intensity are highly dependent on them. The beam radius $w_z$  at a distance $d_{k,v}$ from the VCSEL $v$ to user $k$ is calculated based on the initial beam waist $w_0$ and the Rayleigh range $z_R$. Thus, we have \cite{safi-jlt}
\begin{equation}
w_z(d_{k,v}) = w_0 \sqrt{1 + \left( \frac{d_{k,v}}{z_R} \right)^2}.
\label{beam waist at distance d}
\end{equation}
Furthermore, the intensity of the optical beam at the position of the user $k$ from VCSEL $v$ is described using the Gaussian beam profile, which depends on the radial distance $ r_0^{k,v}$ and the beam radius $w_z(d_{k,v})$. Here, we neglect the effect of side lobes and focus primarily on the main lobes of the Gaussian beam \cite{safi-jlt}. Subsequently, the irradiance $Z_{k,v}$ can be derived as
\begin{equation}
Z_{k,v} = \frac{2}{\pi \big[w_z(d_{k,v})\big]^2} \exp\left( -\frac{2 (r_0^{k,v})^2}{\big[w_z(d_{k,v})\big]^2} \right).
\label{Intensity}
\end{equation}
Finally, the channel gain between VCSEL $v$ and user $k$ is obtained as \cite{H. Kazemi}
\begin{equation}
H_{k,v} = Z_{k,v} \cdot A_{\text{PD}} \cdot G,
\label{Channel}
\end{equation}
where $A_{\text{PD}}$ denotes the PD area, and $G = \frac{n^2}{\sin^2(\theta_r)}$ is the gain of optical lens with $n = 1.5 $ being the refractive index and $ \theta_r $ the half-angle of the receiver's field-of-view (FOV).
\begin{figure}[ht]
	\centering
	\includegraphics[width=0.75\linewidth]{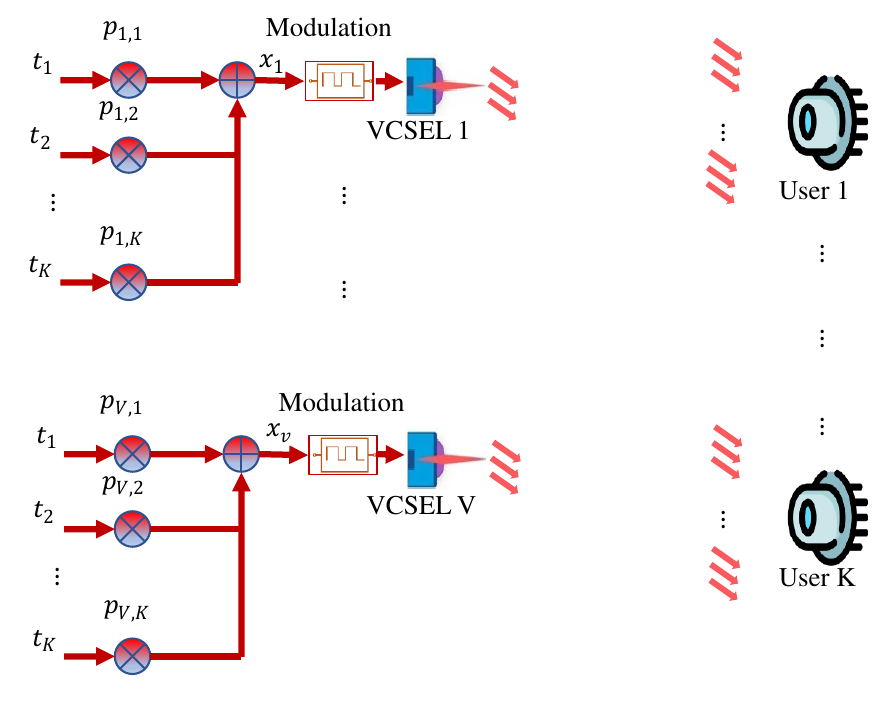}
	\caption{Schematic diagram of a cooperative multiuser broadcast VCSEL-based OWC system.}
	\label{fig:sysmodel}
\end{figure}
\subsection{Signal Model}
 OWC signal detection inherently relies on non-negative, real-valued signals. As shown in Fig. \ref{fig:sysmodel}, let $\mathbf{t} = [t_{1}, t_{2}, \ldots, t_{K}] \in \mathbb{R}^{K \times 1}$ denotes the transmitted signals for all users, with the corresponding precoding matrix 
$\mathbf{P} = [\mathbf{p}_{1}, \mathbf{p}_{2}, \ldots, \mathbf{p}_{K}] \in \mathbb{R}^{V \times K}$, where $\mathbf{p}_{k} = [p_{1,k}, p_{2,k}, \ldots,p_{v,k},\ldots, p_{V,k}]^\top \in \mathbb{R}_+^{V \times 1}$. The transmitted signals are normalized as $\mathbb{E}\{t_{k}\} = 0$ and $\mathbb{E}\{(t_{k})^2\} = 1$ \cite{MUI-2}. The real-valued input electric signal at the $v^{\text{th}}$ VCSEL is obtained by summing the weighted symbols for all $K$ users as follows
\begin{equation}
		x_v = \sum_{k=1}^K p_{v,k} \, t_{k}.
		\label{eq:16}
	\end{equation}
Then, the received signal at the $k$-th user is given by
\begin{equation}
		y_k = \underbrace{\mathbf{h}_k^\top \cdot \rho \mathbf{p}_k \, t_k}_{\text{Desired signal}} + \underbrace{\sum_{l \neq k} \mathbf{h}_k^\top \cdot \rho \mathbf{p}_l \, t_l}_{\text{MUI}}  + n_k
		\label{eq:17}
	\end{equation}
where $\mathbf{h}_k \in \mathbb{R}_+^{V \times 1}$ is the channel vector between the $k^{\text{th}}$ user and the VCSELs array, and $\rho$ denotes photodiode responsivity.  The additive noise at the $k^{\text{th}}$ user is also denoted as $n_k$.

As shown in \eqref{eq:16} and \eqref{eq:17}, the fully cooperative VCSEL array functions as a centralized broadcast transmitter, where signals from all VCSELs are superimposed at each user. This inherently creates a MUI scenario that must be carefully managed to maintain link quality. While coordinated precoding is essential to suppress this interference, we also aim to maximize the overall EE of the system under realistic optical communication constraints, including non-negativity and peak power limits. In what follows, we define a formal expression for system EE and formulate a constrained optimization problem to determine the optimal precoding matrix that not only maximizes EE but also ensures a minimum guaranteed data rate for each user.
\section{Energy Efficiency Analysis}
\label{eeanalysis}
We evaluate the system performance using the signal-to-interference-plus-noise ratio (SINR) criterion, which compares the desired signal power to the interference and noise components as follows
\begin{equation}
    \gamma_k =\frac{N_k}{D_k}= \frac{\left| \mathbf{h}_k^\top \cdot \rho\mathbf{p}_k \,  \right|^2}{\sum_{l \neq k} \left| \mathbf{h}_k^\top \cdot \rho\mathbf{p}_l \,  \right|^2 + n_k}.
    \label{eq:23}
\end{equation}
Nevertheless, given the real-valued and non-negative nature of signals in OWC, the classic Shannon capacity formula, predicated on the assumption of Gaussian input, is not applicable \cite{LiFi Capacity}. Because a closed-form channel capacity is unavailable, we approximate achievable data rates using tight bounds. Specifically, we consider a lower bound of the achievable rate as described in \cite{LiFi Capacity} as follows
\begin{equation}
		R_{k} = \frac{1}{2} \text{BW}_{k} \log_2 \left(1 + \frac{e}{2\pi} \gamma_{k}\right) \text{ bits/sec},
		\label{eq:241}
	\end{equation}
where $\text{BW}_{k}$ is the user bandwidth and $e$ is the Neper number. In a VCSEL-based system, the total power consumption is
\begin{equation}
	P_{\text{VCSEL}} = P_{\text{DC}} + P_{\text{AC}},
	\label{eq:25}
\end{equation}
where \( P_{\text{DC}} \) is the constant bias power and \( P_{\text{AC}} \) is the dynamic power for data transmission, given by \cite{AC power LiFi}
\begin{equation}
	P_{\text{AC}} = \frac{1}{\eta^{\text{VCSEL}}} \sum_{k=1}^K \left\| \boldsymbol{p}_{k} \right\|^2.
	\label{eq:26}
\end{equation}
Here, parameter ${\eta^{\text{VCSEL}}}$ denotes the the power amplifier efficiency \cite{AC power LiFi}. While \( P_{\text{DC}} \) contributes to communication-related energy use, it is independent of the precoding design. Therefore, our optimization focuses on minimizing \( P_{\text{AC}} \), which reflects the effect of signal transmission strategies on energy efficiency. Thus, in the following, we assume $ P_{\text{VCSEL}} = P_{\text{AC}} $. Lastly, to maximize EE  in the considered setup, we define it as the ratio of the achievable rate to the total power consumption \cite{EE}, which is expressed as
\begin{equation}
		\text{EE}^{\text{VCSEL}} = \frac{\sum_{k=1}^K R_{k}}{P_{\text{VCSEL}}}.
		\label{eq:27}
	\end{equation}

\section{Problem formulation and its solution}
\label{III}
In this section, we formulate the EE maximization problem in the considered VCSEL-based OWC system. This optimization enables scalable, sustainable operation in dense deployments by effectively managing MUI and ensuring reliable, high-performance connectivity across the entire coverage area. The EE is maximized by optimizing the precoding vectors $\boldsymbol{p}$ of the users subject to system constraints as follows	
    \begin{equation}
    \text{P1:} \quad {\text{arg}\max}_{\boldsymbol{p}} \, \text{EE}^{\text{VCSEL}}
    \label{eq:281}
\end{equation}
\begin{equation}\small
    \begin{aligned}
        &\text{subject to:} \\
        & \text{C1:} \quad R_k \geq R_k^{\text{min}}, \quad \forall k, \\
        & \text{C2:} \quad P_{\text{VCSEL}} \leq P_{\text{max}}, \\
        & \text{C3:} \quad  p_{v,k} \geq 0, \quad \forall v, k.
    \end{aligned}
    \label{eq:29}
\end{equation}
As detailed in \eqref{eq:29}, system constraints include C1 for maintaining a minimum achievable rate per user, C2 for limiting the transmit optical power of the VCSEL array to comply with eye safety standards, and C3 for ensuring the non-negativity of all transmit precoding coefficients in accordance with the IM/DD requirement. To fulfill eye safety regulations, the VCSELs should operate within the maximum permissible exposure (MPE) defined by the IEC standards \cite{safetypaper}. Therefore, the following conditions should be met
\begin{equation}
P_{\text{max}} \le \frac{1}{\zeta}\pi r_p^2 \times \text{MPE},
\label{safe}
\end{equation}
where $r_p$ is the radius of the pupil. Also in \eqref{safe}, $\zeta$ is obtained as
\begin{equation}\small
\zeta = 1 - \exp\left(-\left(\frac{2r_p^2}{w_z^2(d_\text{haz})}\right)\right),
\label{safe2}
\end{equation}
where $d_\text{haz}$ is defined as the location at which the ratio of the exposure
level to the MPE value is at its maximum.

Before solving Problem $\text{P1}$, we obtain a feasible initialization of the precoding vectors by solving a small convex 
max–min problem that maximizes the minimum user data rate, subject to the transmit power 
constraint (C2), the non-negativity of the precoding coefficients (C3), and convexified rate constraints. This provides a fair and feasible starting point for the iterative algorithm. Given \eqref{eq:27} and \eqref{eq:281}, the problem $\text{P1}$ is a fractional programming problem, where the numerator represents the sum rate, and the denominator includes power consumption, both of which depend on the precoding vectors. This dependency introduces non-convexity into the problem. Additionally, the data rate constraint is also non-convex due to its reliance on the precoding vectors, further increasing the complexity of the optimization. 

To tackle this, we first apply Dinkelbach's approach \cite{Dinkelbach} to transform the non-convex objective function in (\ref{eq:281}) into an equivalent trackable form. Then, we employ the IA technique \cite{IA Framework} to iteratively approximate the non-convex functions and make the problem more manageable. Accordingly, using Dinkelbach's approach, which converts the fractional objective into a difference of two terms, the non-convex problem is transformed as follows
\begin{equation}
    {\text{arg}\max}_{\boldsymbol{p}} \, \left( \sum_{k=1}^K R_{k} - \psi \, P_{\text{VCSEL}} \right),
    \label{eeproblem1}
\end{equation}
where, $\psi$ is a given variable, which will be iteratively updated at each iteration \textit{t} as $\psi^{(t)}= \frac{\sum_{k=1}^K R_{k}^{(t)}}{P_{\text{VCSEL}}^{(t)}}$. This reformulation simplifies the problem, enabling the use of convex optimization techniques like the IA framework to address the non-convexity in both the objective and constraints. Nevertheless, the objective function in \eqref{eeproblem1} and constraint C1 in \eqref{eq:29}  remain non-convex due to the non-convex rate expression $R_k$. Thus, we utilize the IA technique to convexify the rate expression $R_k$ as a function of $\boldsymbol{p}_k$. Let equivalently rewrite $R_k$ in \eqref{eq:241} as 
\begin{equation}
    R_k = \frac{1}{2  {\ln 2}} \text{BW}_k \underbrace{\ln \left( 1 + \frac{e}{2\pi} \gamma_k \right)}_{X_k}
    \label{eq:24}.
\end{equation}
Subsequently, based on \cite[(20)]{Convexify}, a concave approximation of $X_{k}$ can be computed as
\begin{align}
		X_{k} &\geq  X_{k}^{(t)} - \gamma_{k}^{(t)} 
		+ \alpha_{k}-\beta_{k}, \nonumber \\
		&\quad  \overset{\Delta}{=} \overline {X_{k}}
		\label{eq:30}
	\end{align}
where
\begin{align}
	\footnotesize
	&\alpha_{k}=\frac{2e \Re \left[ \left(  \mathbf{h}_{k}^\top \mathbf{p}_{k}^{(t)} \right)^\top  \mathbf{h}_{k}^\top \mathbf{p}_{k} \right]}{2\pi D_{k}^{(t)}}, \nonumber\\
	&\beta_{k}=\frac{\left( eN_{k}^{(t)} \right)}{2\pi D_{k}^{(t)} \left( 2 \pi D_{k}^{(t)} + \left( eN_{k}^{(t)} \right) \right)} \left(  \left(eN_{k}\right) + 2 \pi D_{k}  \right).
\end{align}
    At the $t^{\text{th}}$ iteration, the variables $\gamma_{k}^{(t)}$, $X_{k}^{(t)}$, $N_{k}^{(t)}$, and $D_{k}^{(t)}$ represent the corresponding values of $\gamma_{k}$, $X_{k}$, $N_{k}$, and $D_{k}$, which are constrained within the trust region defined by
\begin{equation}
	\label{llll}
		e \times \left| \mathbf{h}_{k}^\top \mathbf{p}_{k} \right|^2 \geq 0,
	\end{equation}
    where $\times$ denote the multiplication symbol. Also, in \eqref{llll}, $\left| (\mathbf{h}_{k})^\top \mathbf{p}_{k} \right|^2$ can be approximated by the first order Taylor approximation and the equation can be reformulated as
    \begin{align}
		& e \times \Big\{2 \Re \left[ \left(  \mathbf{h}_{k}^\top \mathbf{p}_{k}^{(t)} \right)^\top  \mathbf{h}_{k}^\top \mathbf{p}_{k} \right]- \left|  \mathbf{h}_{k}^\top \mathbf{p}_{k}^{(t)} \right|^2 \Big\} \geq 0 .
	\end{align}

    Now, the concave rate expression $\overline {R_{k}}$ can be represented as
    \begin{equation}
    \overline {R_{k}} = \frac{1}{2 . {\ln 2}} \text{BW}_k \overline {X_{k}}.
    \label{eq:24}
\end{equation} 

Eventually,  the non-convex optimization problem P1  presented in \eqref{eq:281}  can be reformulated as the following convex optimization problem
 \begin{equation}
    \text{P2:}  \quad   {\text{arg}\max}_{\boldsymbol{p}} \, \left( \sum_{k=1}^K \overline {R_{k}} - \psi \, P_{\text{VCSEL}} \right),
    \label{eq:282}
\end{equation}
Subject to:
\begin{equation}
	\small
	\begin{aligned}
		& \text{C1:} \quad \overline {R_{k}} \geq R_k^{\text{min}}, \quad \forall k, \\
		& \text{C2:} \quad P_{\text{VCSEL}} \leq P_{\text{max}}, \\
		& \text{C3:} \quad  p^{(t)}_{v,k} \geq 0, \quad \forall v, k.\\
		& \text{C4:} \quad  e\Big\{2 \Re \left[ \left(  \mathbf{h}_{k}^\top \mathbf{p}_{k}^{(t)} \right)^\top  \mathbf{h}_{k}^\top \mathbf{p}_{k} \right]- \left|  \mathbf{h}_{k}^\top \mathbf{p}_{k} \right|^2 \Big\} \geq 0 , \quad \forall k.
	\end{aligned}
	\label{eq:292}
\end{equation}
We can now iteratively solve this convexified optimization problem using optimization tools such as MATLAB CVX. 


\section{Performance Evaluation}
\label{IV}
\begin{figure}
	\centering
	\includegraphics[width=0.6\linewidth]{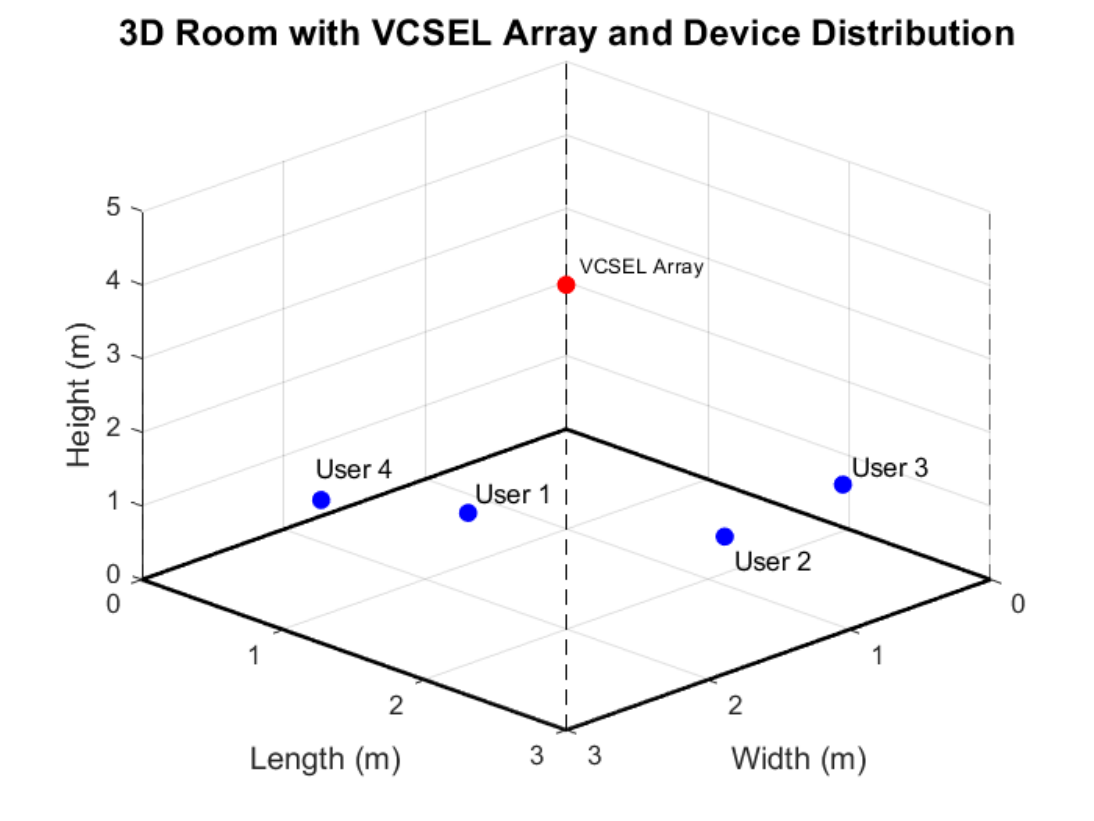}
	\caption{Simulated environment (the VCSEL array and the randomly distributed users).}
	\label{FIG:7}
\end{figure}

We simulate a $3\text{m} \times 3\text{m} \times 5\text{m}$ indoor environment with VCSEL arrays arranged in $4\times4$, $5\times5$, $6\times6$, $7\times7$, and $8\times8$ grid configurations, mounted at a height of $4\text{m}$ \cite{VCSEL3}. Four user are randomly located at discrete heights of $0.5\text{m}$, $1\text{m}$, $1.5\text{m}$, and $2\text{m}$. Each configuration is evaluated independently to assess the effect of array density on EE. Fig. \ref{FIG:7} depicts the VCSEL deployment and user distribution\footnote{Although for the sake of simplicity in the simulations we assumed this arrangement for the user location, our proposed approach in this paper can be readily applied to an arbitrary number of users with arbitrary locations.}. Simulation parameters are summarized in Table \ref{tbl2}. \FloatBarrier

\begin{table}[h]
	\centering
	\caption{Simulation Parameters}
	\begin{tabular}{@{}lc|lc@{}}
		\hline
		\textbf{Parameter} & \textbf{Value} & \textbf{Parameter} & \textbf{Value} \\
		\hline
		Beam waist ($w_0$) & 0.45\,$\mu$m & Detector responsivity ($\rho$) & 1 A/W \\
		Photodetector area ($A_\text{PD}$) & 1\,$\textrm{cm}^2$ & Pitch Distance ($\textrm{pi}_x,\textrm{pi}_y$) & 10\,$\mu$m \\
		Lens ref. index ($n$) & 1.5 &  Bandwidth (BW) & 2\,GHz \\
		VCSEL wavelength ($\lambda$) & 950\,nm & & \\
		\hline
	\end{tabular}
	\label{tbl2}
\end{table}

\begin{figure}
	\centering
	\includegraphics[width=0.6\linewidth]{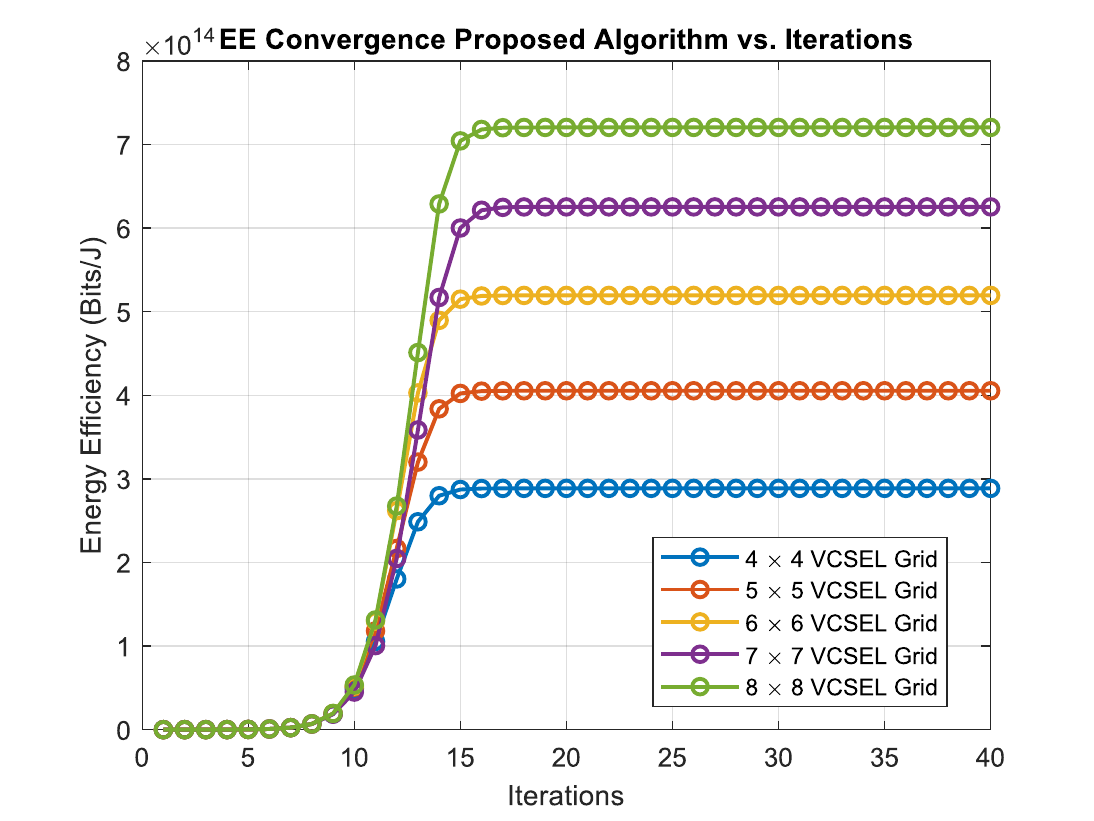}
	\caption{EE convergence of the proposed algorithm for different VCSEL array grid sizes.}
	\label{FIG:8}
\end{figure}

Fig. \ref{FIG:8} shows the EE convergence of the proposed algorithm for different VCSEL array grid sizes. As the grid size increases, EE rises rapidly initially, then stabilizes at a steady value. This indicates that the algorithm effectively optimizes EE as the number of lasers in the array grows. The convergence trend demonstrates that our approach remains robust in dense configurations, a key requirement for large-scale OWC networks. As illustrated in Fig. \ref{FIG:8}, the proposed algorithm converges within a small number of iterations. 
Since the initialization is obtained from a convex max–min problem, the method starts from a feasible and fair point, leading to stable convergence behavior. Given the non-convexity of P1, the algorithm converges to a locally optimal solution, which is standard for SCA-based methods.

\begin{figure} [h]
    \centering
    \includegraphics[width=0.6\linewidth]{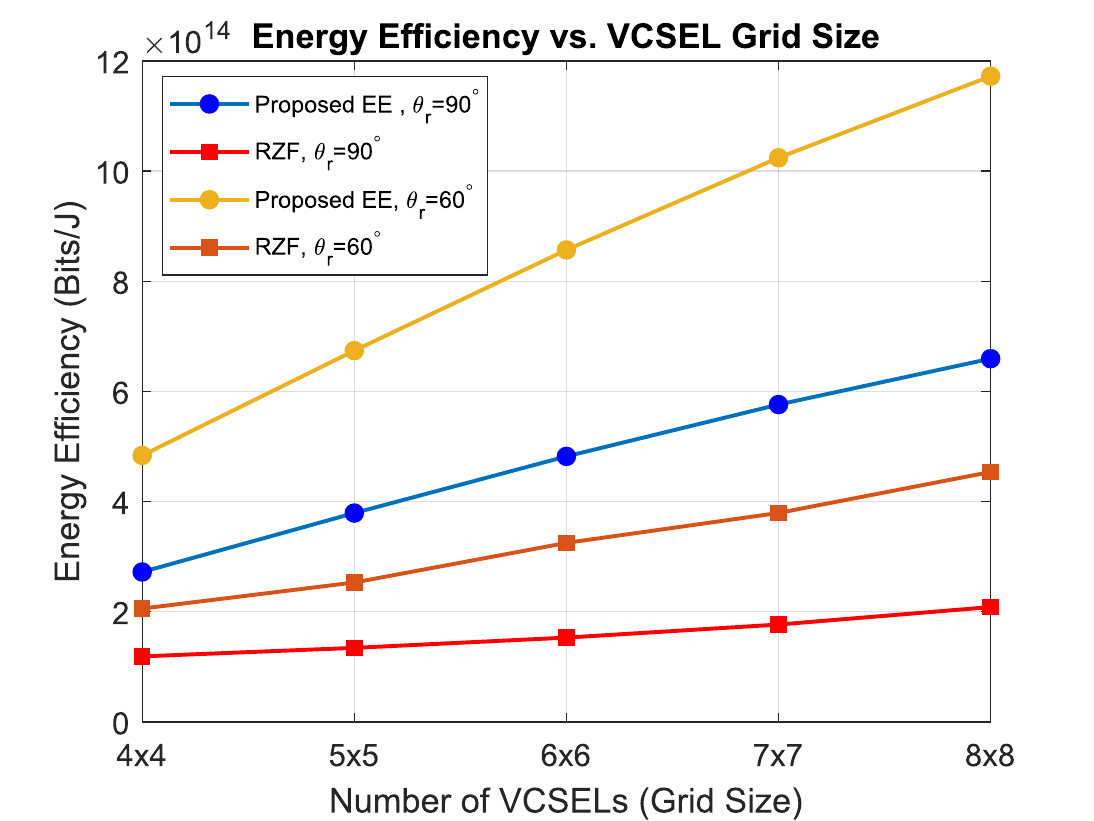}
    \caption{EE performance comparison between the proposed algorithm and the RZF precoding benchmark (under various VCSEL array grid configurations and receiver FoV)}
    \label{EEvsRZF}
\end{figure}

Fig. \ref{EEvsRZF} compares the EE performance of the proposed precoding algorithm with that of the RZF benchmark under varying VCSEL array grid sizes and receiver FoV settings. The results clearly demonstrate that our method consistently outperforms RZF across all configurations. This improvement stems from our joint optimization of the precoding matrix, which effectively balances interference suppression and energy usage across the entire cooperative VCSEL array, unlike RZF, which often incurs excessive power consumption in dense deployments due to poor channel conditioning. As expected, EE increases with larger array grids for both schemes, owing to greater spatial degrees of freedom. On the other hand, increasing the receiver FoV leads to reduced EE for both approaches. A wider FoV captures more interfering signals due to its broader angular acceptance, which lowers the SINR and thus degrades achievable data rates and EE. While narrower FoVs improve SINR and EE by limiting interference, they also pose practical challenges in terms of precise beam steering and user alignment.

Finally, we assess the computational complexity of the proposed method, which combines Dinkelbach’s and IA iterations. Each Dinkelbach iteration solves a convex problem with complexity $\mathcal{O}((V + K)^3)$, while each IA iteration has complexity $\mathcal{O}(V^3 + K^2)$, where $V$ and $K$ denote the number of VCSELs and users, respectively. Letting $I_D$ and $I_{\text{IA}}$ represent the iteration counts, the overall complexity is
\begin{equation}
	\mathcal{O}\left(I_D \cdot (V + K)^3 + I_\text{IA} \cdot \left(V^3 + K^2\right)\right),
\end{equation}
which remains tractable for moderately dense deployments. For comparison, the RZF benchmark requires computing the regularized Gram matrix inverse $\left(HH^\top + \alpha I\right)^{-1}$, with an overall complexity of $\mathcal{O}(K^2 V + K^3 + VK^2)$. While RZF offers a lower complexity focused solely on interference suppression, our method explicitly targets energy efficiency and achieves substantial performance gains with moderate additional computational cost.

\section{Conclusion}
\label{V}

We presented an energy-efficient precoding framework for dense VCSEL-based OWC systems under a cooperative broadcast model. By leveraging full-array transmission and optimizing the precoding matrix via Dinkelbach’s method and inner approximation, our approach effectively manages MUI and maximizes EE. The obtained results demonstrate substantial improvements in the EE of the proposed precoding algorithm over conventional RZF, with manageable complexity suitable for practical deployment.


%



\section*{Acknowledgment}
This work is a contribution by Project REASON, a UK Government funded project under the Future Open Networks Research Challenge (FONRC) sponsored by the Department of Science Innovation and Technology (DSIT). The authors also acknowledge support by the Engineering and Physical Sciences Research Council (EPSRC) under grants EP/X04047X/1 - EP/Y037243/1 ``Platform for Driving Ultimate Connectivity (TITAN).''

\ifCLASSOPTIONcaptionsoff
  \newpage
\fi






\balance

\end{document}